\documentclass[12pt]{article} 
\usepackage{pdproc,epsfig}  
 
\begin{document} 
 
\renewcommand{\thefootnote}{\alph{footnote}} 
   
\title{ 
 PHYSICS AT A NEW\\ 
 FERMILAB PROTON DRIVER} 
 
\author{ Steve Geer 
\footnote{Presented at the III International Workshop on Neutrino Oscillations in Venice, Feb 7-10, 2006}
} 
 
\address{Fermi National Laboratory, PO Box 500,
  \\ 
 Batavia, IL 60510, U.S.A.\\ 
 {\rm E-mail: sgeer@fnal.gov}}

\abstract{
In 2004, motivated by the recent exciting developments in neutrino physics,
 the Fermilab Long Range Planning Committee identified a new high 
intensity Proton Driver as an attractive option for the future. At the end 
of 2004 the APS ``Study on the Physics of Neutrinos'' concluded that 
the future U.S. neutrino program should have, as one of its components, 
``A proton driver in the megawatt class or above and neutrino superbeam 
with an appropriate very large detector capable of observing CP violation 
and measuring the neutrino mass-squared differences and mixing parameters 
with high precision''. The presently proposed Fermilab Proton Driver 
is designed to accomplish these goals, and is based on, and would help 
develop, Linear Collider technology. In this paper the Proton Driver 
parameters are summarized, and the potential physics program is described.
} 
    
\normalsize\baselineskip=15pt 
 
\section{Introduction}

The interest in a new Fermilab Proton Driver is motivated by the exciting 
discoveries that have been made in the neutrino sector. 
In the last few years solar, atmospheric, and reactor neutrino experiments 
have revolutionized our understanding of the nature of neutrinos. We now 
know that neutrinos produced in a given flavor eigenstate can transform 
themselves into neutrinos of a different flavor as they propagate over 
macroscopic distances. This means that, like quarks, neutrinos have a 
non-zero mass, the flavor eigenstates are different from the mass eigenstates, 
and hence neutrinos mix. Understanding neutrino properties is important 
because neutrinos are the most common matter particles in the universe. 
In number, they exceed the constituents of ordinary matter (electrons, 
protons, neutrons) by a factor of ten billion. However, we have incomplete 
knowledge of their properties since {\it we do not know the spectrum of 
neutrino masses, and we have only partial knowledge of the mixing among 
the three known neutrino flavor eigenstates}.  Furthermore, it is possible 
that the simplest three-flavor mixing scheme is not the whole story, and 
that a complete understanding of neutrino properties will require a more 
complicated framework. In addition to determining the parameters that 
describe the neutrino sector, the three-flavor mixing framework must be tested.

To identify the best ways to address the most important open neutrino 
questions, and to determine an effective, fruitful U.S. role within a 
global experimental neutrino program, the American Physical Society's 
Divisions of Nuclear Physics and Particles and Fields, together with the 
Divisions of Astrophysics and the Physics of Beams, have recently conducted 
a ``Study on the Physics of Neutrinos''. This study 
recommended~\cite{the-neutrino-matrix}
{\it ``... as a high priority, a comprehensive U.S. program to 
complete our understanding of neutrino mixing, to determine the character 
of the neutrino mass spectrum, and to search for CP violation among 
neutrinos'' }, and identified, as a key ingredient of the 
future program,  
{\it ``A proton driver in the megawatt class or above and neutrino superbeam 
with an appropriate very large detector capable of observing CP violation 
and measuring the neutrino mass-squared differences and mixing parameters 
with high precision.'' } The proposed Fermilab Proton Driver would, 
together with a suitable new detector,  
fulfill this need by providing a 2~megawatt proton beam at Main Injector 
(MI) energies for the future NuMI~\cite{numi} program.
   
\begin{table}[h] 
\caption{\label{tab:params} Parameters for the present Fermilab Proton Source compared 
with the corresponding parameters for the proposed Fermilab Proton 
Driver. }
  \small 
  \begin{tabular}{||l|c|c||}\hline\hline 
  {} &{} &{} \\ 
  Parameter & Present & Proton Driver \\ 
  {} &{} &{} \\ 
  \hline 
  {} &{} &{} \\
{\bf Linac} &{} &{}\\
Pulse Frequency & 5 Hz & 10 Hz\\
Kinetic Energy & 400 MeV & 8 GeV \\
Peak Current & 40 mA & 28 mA \\
Pulse Length & 25 $\mu$s & 1 ms \\
  {} &{} &{} \\
  \hline
  {} &{} &{} \\
{\bf Booster} &{} &{}\\
Cycle Frequency & 15 Hz &{} \\
Extraction Kinetic Energy & 8 GeV &{} \\
Proton per cycle & $5 \times 10^{12}$ &{} \\
Proton per hour & $9 \times 10^{16}$ (5 Hz) &{} \\
8 GeV Beam Power & 33 KW (5 Hz) & 2 MW\\
  {} &{} &{} \\
  \hline
  {} &{} &{} \\
{\bf Main Injector} &{} &{}\\
Kinetic Energy & 120 GeV &{} \\
Protons per cycle & $3 \times 10^{13}$ & $1.5 \times 10^{14}$ \\
Fill time & 0.4 sec & 0.1 sec \\
Ramp time & 1.47 sec & 1.4 sec\\
Cycle time & 1.87 sec & 1.5 sec\\
Protons per hour & $5.8 \times 10^{16}$ & $3.5 \times 10^{17}$\\
  {} &{} &{} \\
  \hline\hline 
\end{tabular}  
\end{table}

\section{Proton Driver Parameters}

The design that is currently favored~\cite{foster,pd-website,kephart} 
consists of an 8~GeV $H^-$ superconducting (SC) Linac that utilizes 
International Linear Collider (ILC) technology. 
In Table~\ref{tab:params} the parameters describing the 
performance of the present Fermilab Proton Source are compared with 
the corresponding parameters for the Fermilab Proton Driver.
The new linac would  
produce a 0.5~megawatt beam which could be upgraded to 2~megawatts. 
A small fraction of the 8~GeV beam would be used to 
fill the Fermilab Main Injector (MI) with the maximum number of protons 
that, with some modest improvements, it can accelerate. 
This would yield a 2~megawatt MI beam at an 
energy that can be chosen to be anywhere within the range 40~GeV to 120~GeV.
Hence the upgraded proton source would simultaneously deliver two beams: 
a 2~megawatt beam at MI energies, and an  $\sim 0.5 - 2$~megawatt beam 
at 8~GeV.
\begin{figure}[b!]
\begin{center}
\scalebox{.65}{\rotatebox{270}
{\includegraphics*[bb=138 61 532 721,clip]{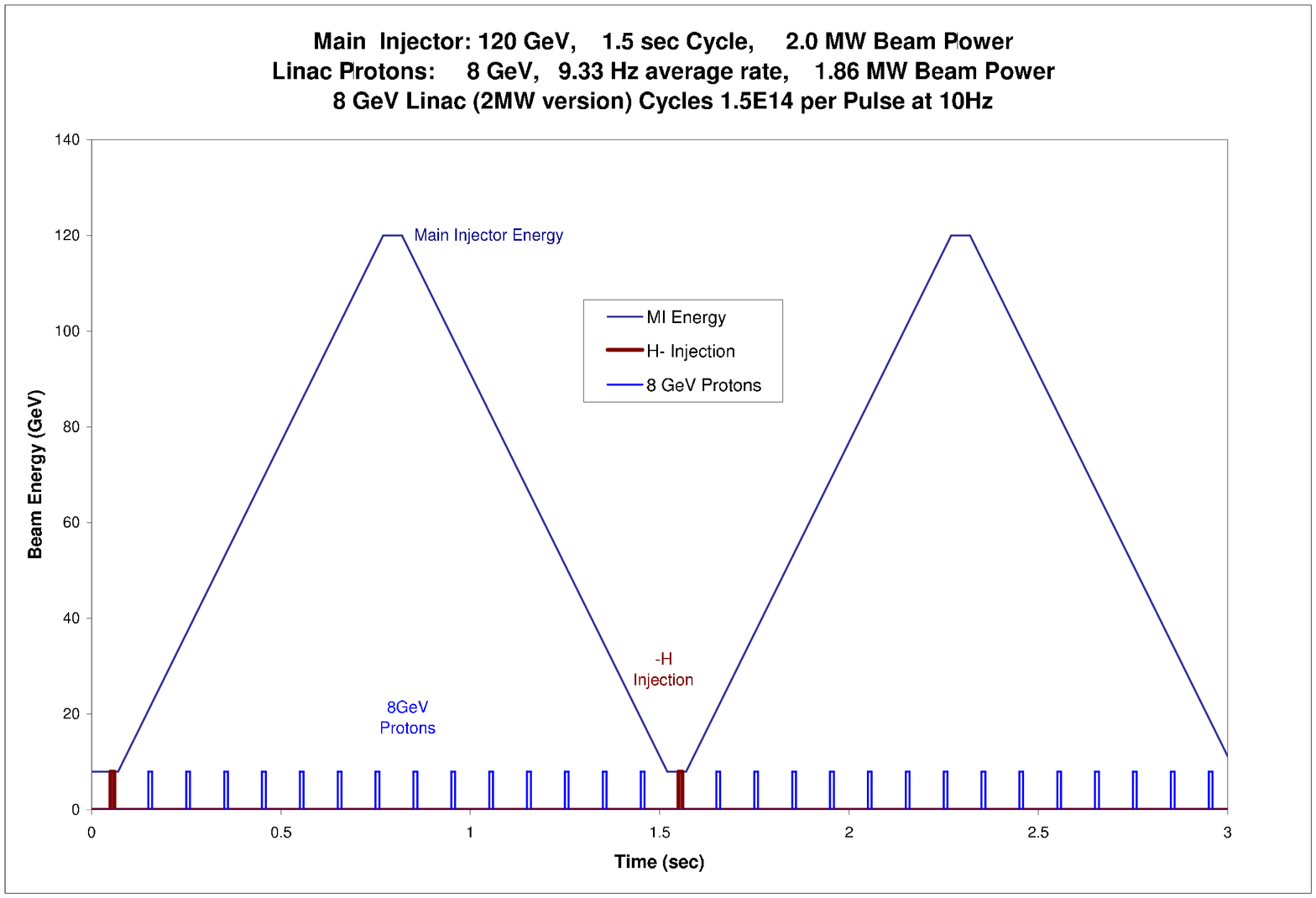}}}
\end{center}
   \caption{\label{fig:MIcycle} 
Proton Driver bunch structure and the Main Injector cycle.
}
\end{figure}

To illustrate this the cycle structure is shown in Fig.~\ref{fig:MIcycle}. 
The MI would receive one pulse from the Linac 
every 1.5~sec. Note that the MI fill time is very short ($<1$~ms).  
The MI cycle time is dominated by the time to ramp up to 120~GeV and ramp down 
to 8~GeV. The 14 Linac pulses that are available, while the MI is ramping 
and at flat top, would provide beam for an 8~GeV program. Improvements in the 
MI ramping time might eventually enable more of the 8~GeV Linac beam to be 
accelerated in the MI, yielding beam powers exceeding 2~megawatts.

\section{Neutrino Oscillations}

To understand the neutrino oscillation physics reach at the Fermilab 
Proton Driver we first introduce the three-flavor mixing parameters, 
and then discuss event rates and discovery potential.

\subsection{Three-Flavor Mixing Parameters}

There are three known neutrino flavor eigenstates 
$\nu_\alpha = (\nu_e, ~\nu_\mu, ~\nu_\tau)$. Since transitions have been 
observed between the flavor eigenstates we now know that neutrinos have 
non-zero masses, and that there is mixing between the flavor eigenstates.
The mass eigenstates $\nu_i = (\nu_1, ~\nu_2, ~\nu_3)$ with masses 
$m_i = (m_1, ~m_2, ~m_3)$ are related to the flavor eigenstates by 
a $3 \times 3$ unitary mixing matrix $U^\nu$~\cite{mns}, 
\begin{equation}
  |\nu_\alpha\rangle = \sum_i ( U^\nu_{\alpha i} )^* |\nu_i\rangle
\label{mix}
\end{equation}
Four numbers are needed to specify all of the matrix elements, 
namely three mixing 
angles ($\theta_{12}, \theta_{23}, \theta_{13}$) and one complex 
phase ($\delta$). In terms of these parameters:  $ U^\nu =$
\begin{equation}
\left( \begin{array}{ccc}
  c_{13} c_{12}       & c_{13} s_{12}  & s_{13} e^{-i\delta} \\
\\
-c_{23} s_{12}
& c_{23} c_{12}
& c_{13} s_{23} \\
-s_{13} s_{23} c_{12} e^{i\delta}
& -s_{13} s_{23} s_{12} e^{i\delta}
& \\
\\
    s_{23} s_{12}
& -s_{23} c_{12}
& c_{13} c_{23}\\
   -s_{13} c_{23} c_{12} e^{i\delta}
& -s_{13} c_{23} s_{12} e^{i\delta}
& 
\end{array} \right) \,
\label{mns}
\end{equation}
where $c_{jk} \equiv \cos\theta_{jk}$ and $s_{jk} \equiv \sin\theta_{jk}$. 
Neutrino oscillation measurements have already provided some knowledge 
of $U^\nu$, which is approximately given by:
\begin{equation}
U^\nu =
\left( \begin{array}{ccc}
  0.8  & 0.5  & ? \\
  0.4  & 0.6  & 0.7 \\
  0.4  & 0.6  & 0.7 \\
\end{array} \right) \,
\label{mnsnumbers}
\end{equation}
We have limited knowledge of the (1,3)-element of the mixing matrix. This 
matrix element is parametrized by $s_{13} e^{-i\delta}$. We have only an upper 
limit on $\theta_{13}$ and no knowledge of $\delta$. 
Note that $\theta_{13}$ and $\delta$ are particularly important because if 
$\theta_{13}$ and $\sin \delta$ are non-zero there will be CP violation in 
the neutrino sector.

\begin{figure}[t]
\begin{center}
\includegraphics[width=0.9\textwidth]{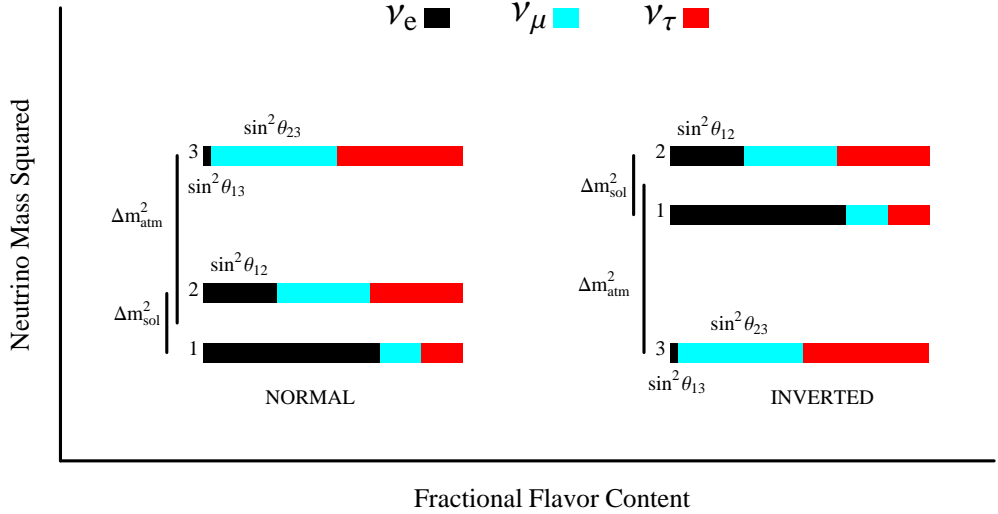}
\end{center}
\caption{\label{hierfig}
The two possible arrangements of the masses of the three known neutrinos, 
based on neutrino oscillation measurements. The spectrum on the left 
corresponds to the {\it Normal Hierarchy} and has $\Delta m^2_{32} > 0$. 
The spectrum on the right corresponds to the {\it Inverted Hierarchy} and 
has $\Delta m^2_{32} < 0$.
The $\nu_e$ fraction of each mass eigenstate is indicated by the black solid
region. The $\nu_\mu$ and $\nu_\tau$ fractions are indicated by
the blue (red) regions respectively.
The $\nu_e$ fraction in the
mass eigenstate labeled ``3'' has been set to the CHOOZ bound. Figure from 
Ref.~7.
}
\end{figure}
Neutrino oscillations are driven by the splittings between the neutrino 
mass eigenstates. 
It is useful to define the differences between the squares of the masses of 
the mass eigenstates $\Delta m^2_{ij} \equiv m^2_i-m^2_j$. 
The probability that a neutrino of energy $E$ and initial flavor $\alpha$ will 
``oscillate'' into a neutrino of flavor $\beta$ is given by  
$P_{\alpha \beta} \equiv P(\nu_\alpha \rightarrow \nu_\beta) =
\left| \langle \nu_\beta | \exp( - i \mathcal{H} t ) | \nu_\alpha \rangle \right|^2$,
which in vacuum is given by 
\begin{equation}
P_{\alpha \beta}  =  
\sum\limits_{j=1}^3 \, \sum\limits_{k=1}^3 U_{\alpha j} U_{\alpha k}^* U_{\beta j}^* U_{\beta k} \, 
\exp \left( - i \frac{\Delta m_{kj}^2}{2 E} t \right)
\label{equ:oscgeneral}
\end{equation}
If neutrinos of energy $E$ travel a distance $L$ then a 
non-zero $\Delta m^2_{ij}$ will result in neutrino flavor 
oscillations that have maxima at given values of $L/E$, and oscillation 
amplitudes that are determined by the matrix elements $U^\nu_{\alpha i}$, and 
hence by $\theta_{12}, \theta_{23}, \theta_{13}$, and $\delta$. 

\begin{table}[h] 
\caption{\label{table:1} 
Signal and background $\nu_\mu \rightarrow \nu_e$ event rates
for values of $\theta_{13}$ that are just below the present upper limit 
and an order of magnitude below the upper limit. The rates are for 
the normal mass hierarchy and $\delta = 0$. The numbers for each 
experiment correspond to 5~years of running with the nominal beam intensities.
 } 
  \small 
  \begin{tabular}{||l|c|c|c||}\hline\hline 
  {} &{} &{} &{} \\
Experiment  & Signal & Signal & Background \\
          {}& $\sin^2 2\theta_{13}=0.1$ & $\sin^2 2\theta_{13}=0.01$ & \\
  {} &{} &{} &{} \\
\hline
  {} &{} &{} &{} \\
MINOS  & 49.1 & 6.7 & 108\\
ICARUS & 31.8 & 4.5 & 69.1\\
OPERA  & 11.2 & 1.6 & 28.3\\
T2K    & 132  & 16.9 & 22.7\\
NO$\nu$A& 186&23.0&19.7\\
  {} &{} &{} &{} \\
\hline
  {} &{} &{} &{} \\
NO$\nu$A$+$FPD&716&88.6&75.6\\
NuFACT (neutrinos)&29752&4071&44.9\\
NuFACT (antineutrinos)&7737&1116&82.0\\
  {} &{} &{} &{} \\
\hline\hline
\end{tabular}\\
From the calculations of W.~Winter, based on the Globes 
program~\cite{Huber:2002mx}.
\end{table}

\begin{table}[h]
\caption{\label{table:2} 
Signal and background $\nu_\mu \rightarrow \nu_e$ event rates
for $\theta_{13}$ just below the present upper limit 
($\sin^2 2 \theta_{13} = 0.1$). The rates are for 
the normal and inverted mass hierarchies with $\delta = 0$ (no CP violation) 
and $\delta = \pi/2$ (maximal CP violation). The numbers for each 
experiment correspond to 5~years of running with the nominal beam intensities.
 }
  \small
\begin{tabular}{||l|cc|cc|c||}\hline\hline
  {} &{} &{} &{} &{} &{} \\
Experiment       & Normal & Normal & Inverted & Inverted & Back- \\
                 &$\delta=0$&$\delta=\pi/2$&$\delta=0$&$\delta=\pi/2$&ground \\
  {} &{} &{} &{} &{} &{} \\
\hline
  {} &{} &{} &{} &{} &{} \\
T2K    & 132  & 96 & 102& 83& 22.7\\
NO$\nu$A& 186&138&111&85&19.7\\
NO$\nu$A$+$FPD&716&531&430&326&75.6\\
  {} &{} &{} &{} &{} &{} \\
NuFACT ($\nu$)&29752&27449&13060&17562&44.9\\
NuFACT ($\overline{\nu}$)&7737&5942&9336&10251&82.0\\
  {} &{} &{} &{} &{} &{} \\
\hline
\end{tabular}\\
From the calculations of W.~Winter, based on the Globes 
program~\cite{Huber:2002mx}.
\end{table}

The mixing angles tell us about the flavor content of the neutrino mass 
eigenstates. Our knowledge of the $\Delta m^2_{ij}$ and the flavor content 
of the mass eigenstates is summarized in Fig.~\ref{hierfig}. Note that 
there are two possible patterns of neutrino mass. 
This is because the neutrino oscillation experiments to date have been 
sensitive to the magnitude of $\Delta m^2_{32}$, but not its sign. 
The neutrino spectrum shown on the left in Fig.~\ref{hierfig} is called 
the {\it Normal Mass Hierarchy} and corresponds to $\Delta m^2_{32} > 0$.  
The neutrino spectrum shown on the right is called the {\it Inverted Mass 
Hierarchy} and corresponds to $\Delta m^2_{32} < 0$. The reason we don't 
know the sign of $\Delta m^2_{32}$, and hence the neutrino mass hierarchy, 
is that neutrino oscillations in vacuum depend only on the magnitude of 
$\Delta m^2_{32}$. However, in matter the effective parameters describing 
neutrino transitions involving electron-type neutrinos are 
modified~\cite{wolfenstein} in a way that is 
sensitive to the sign of $\Delta m^2_{32}$. An experiment with a sufficiently 
long baseline in matter and an appropriate $L/E$ can therefore determine 
the neutrino mass hierarchy.

\subsection{Event Rates}

To obtain sufficient statistical sensitivity to determine the pattern of 
neutrino masses and search for CP violation over a large region 
of parameter-space will require a new detector with a fiducial mass of tens
of kilotons and a neutrino beam with the highest practical intensity. 
To illustrate this, consider the NuMI event rates in the far detector. 
The present NuMI primary proton beam intensity is roughly $10^{13}$~protons 
per second at 120~GeV, which corresponds to 0.2~megawatts on target. 
These protons are used to make a secondary charged pion beam, which is 
focused into a parallel beam using magnetic horns. The pion beam is 
then allowed to decay whilst propagating down a long decay channel, 
to create a tertiary beam of muon-neutrinos. At the far detector, 
735~km downstream of the target, there are $10^{-5}$ neutrino interactions 
in a 1~kt detector for every $10^{13}$ protons on target. Note that we 
are interested in $\nu_\mu \rightarrow \nu_e$ oscillations, and that 
the present upper limit on $\theta_{13}$ implies that the relevant 
oscillation amplitude is at most $\sim5$\%. Putting these numbers 
together one quickly concludes that we will need proton beam powers 
of one or a few megawatts together with detectors of a few times 
10~kt.

To be explicit, the expected $\nu_\mu \rightarrow \nu_e$ event rates 
for future experiments are listed in Table~\ref{table:1} for two values 
of $\theta_{13}$. Note that signal and background rates for the T2K and 
NO$\nu$A~\cite{nova} experiments are comparable and will at best (for the most 
favorable $\theta_{13}$) yield data samples of only $\sim 100$~events. 
This may be sufficient to pin down $\theta_{13}$, but is at best barely 
sufficient to make the first determination of the mass hierarchy, and is 
inadequate to search for CP violation or make precision measurements of 
the interesting parameters. The Fermilab Proton Driver, together with 
NO$\nu$A, does significantly better. In the longer term the Proton Driver 
could be used to ``drive'' a Neutrino Factory~\cite{nf}, 
which would improve the signal rates by approaching two orders of magnitude. 

Although the signal and background rates for T2K and NO$\nu$A are 
comparable, the two experiments are complementary in the way they 
are sensitive to the oscillation parameters. This is illustrated 
in Table~\ref{table:2} which compares event rates for the two 
mass hierarchies and for CP conserving and CP violating values of $\delta$.
With the Proton Driver, there is a statistically significant 
dependence of event rates on both the mass hierarchy and the phase 
$\delta$. In contrast, the T2K rates are not as sensitive to the 
mass hierarchy. Hence the combination of both experiments provides a  
way to disentangle the parameters.

\begin{figure}[t]
\begin{center}
\includegraphics[width=0.7\textwidth]{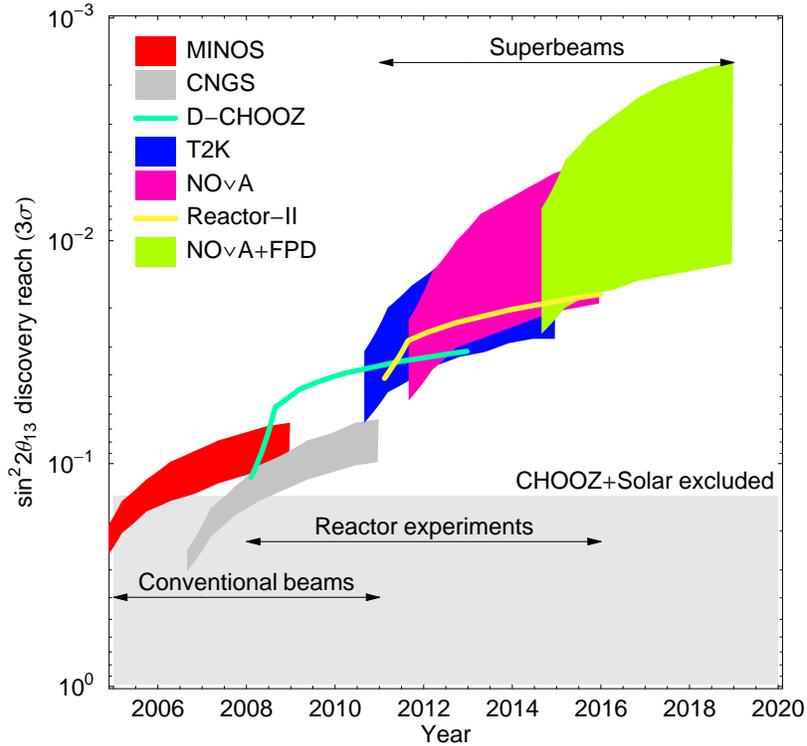}
\end{center}
\caption{\label{fig:timescale_small} Anticipated evolution of the 
$\theta_{13}$ discovery reach. The 3$\sigma$ sensitivities for the 
observation of a non-zero $\sin^2 2\theta_{13}$. The bands reflect the 
dependence on the CP phase $\delta$. The calculations are for a normal mass 
hierarchy and are 
based on the simulations in Ref.~8, 11 
and include 
statistical and systematic uncertainties and parameter correlations. 
All experiments are operated with neutrino running only. 
The starting times of the experiments correspond to those 
stated in the respective LOIs. ReactorII and FPD refer, respectively, to a 
second generation reactor experiment and to the Fermilab Proton Driver.
   }
\end{figure}
\begin{figure}[h]
\begin{center}
\includegraphics[width=0.9\textwidth]{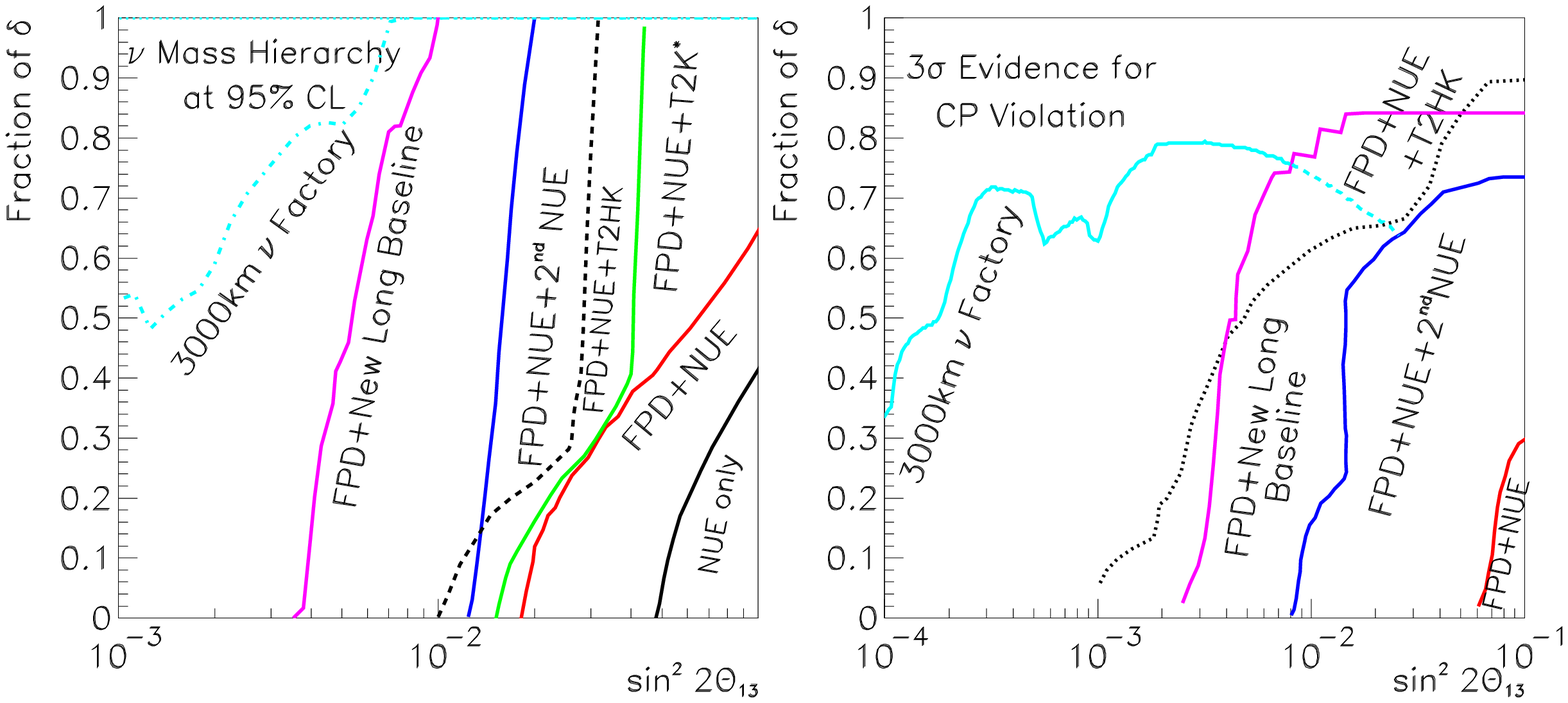}
\end{center}
\caption{\label{fig:reach_all_pd}
Regions of parameter space where the mass hierarchy (left) and CP 
violation (right) can be observed at 95\% CL and at 3$\sigma$, 
respectively. The label NUE refers to the NO$\nu$A experiment, 
and FPD to the Fermilab Proton Driver. 
T2K$^\star$ refers to an upgraded T2K experiment with 
a 4~megawatt primary beam.}
\end{figure}

\subsection{Discovery Potential}

To complete our knowledge of the neutrino mixing matrix and the pattern 
of neutrino masses we must measure $\theta_{13}$ and $\delta$, determine 
the sign of $\Delta m^2_{32}$, and test the three-flavor mixing framework. 
The initial goal for a Fermilab Proton Driver experiment will be to make 
these measurements. How far this physics program can be pursued will depend 
upon the magnitude of the unknown mixing angle $\theta_{13}$.

The anticipated evolution of the $\sin^2 2\theta_{13}$ discovery reach of the 
global neutrino oscillation program is illustrated in 
Fig.~\ref{fig:timescale_small}. 
The sensitivity is expected to improve by about an order of magnitude over the 
next decade. This progress is likely to be accomplished in several steps, 
each yielding a factor of a few increased sensitivity. 
During this first decade the Fermilab program will have contributed to the 
improving global sensitivity with MINOS, followed by NO$\nu$A. The Proton 
Driver would then take NO$\nu$A into the fast lane of the global program.

Although we don't know the value of $\theta_{13}$ we have no reason to suspect 
that it is very small. Hence, any of the experiments on the trajectory shown 
in Fig.~\ref{fig:timescale_small} might establish a finite value for 
$\theta_{13}$. The focus of the experimental program would then change from 
establishing the magnitude of $\theta_{13}$ to measuring the mass hierarchy 
and searching for CP violation. 


The experiments needed to determine the mass hierarchy and discover (or place 
stringent limits on) CP violation will depend upon both $\theta_{13}$ and on 
$\delta$. The fractions of all possible values of $\delta$ for which a discovery 
can be made are shown as a function of $\sin^2 2\theta_{13}$  in 
Fig.~\ref{fig:reach_all_pd} for various experiments. The left panel shows the 
potential for determining the mass hierarchy and the right panel for making a 
first observation of CP violation. Note that without a megawatt-class proton source 
none of the future experiments will be able to make a sensitive search for 
CP violation. The NO$\nu$A experiment (labeled NUE in the figure) can make a 
first determination of the mass hierarchy, but only over a very limited region 
of parameter space. The Fermilab Proton Driver would significantly improve the 
prospects for determining the mass hierarchy, and if $\theta_{13}$ is relatively 
large, would enable the first sensitive search for CP violation. Combining NO$\nu$A 
and T2K results would enable further progress if the T2K experiment was upgraded 
to achieve a factor of a few larger data samples (T2K$^\star$). The mass hierarchy 
could then be determined independent of $\delta$ provided $\sin^2 2\theta_{13}$ 
exceeds about 0.04. Smaller values of $\theta_{13}$ will motivate a much more 
ambitious experimental program which will probably include a Neutrino Factory. 
Larger values of $\theta_{13}$ will still motivate a more ambitious experimental 
program focused on the precision measurements that would put the presently viable
theoretical models under pressure. The second generation of Fermilab Proton Driver 
experiments might, in this case, include a second off-axis detector and/or a new 
longer-baseline beam. Note that, since a Proton Driver can be used to drive 
a Neutrino Factory, the Fermilab Proton Driver offers great flexibility for a 
second generation program independent of the value of $\theta_{13}$.

\subsection{Muon Physics}

A Fermilab Proton Driver could support, in addition to a program of neutrino 
oscillation experiments, a broader program of experiments using neutrino, 
proton, pion, muon, kaon, and neutron beams. Recent studies have identified 
a muon-based physics program as particularly promising. The focus 
of the muon program would be a search for, and perhaps measurement of, 
lepton-flavor- and CP- violation in the charged lepton sector, complementing 
the corresponding measurements made by 
neutrino oscillation experiments in the neutral lepton sector. 

Low energy high precision muon experiments require high intensity beams. 
Since most of the 8~GeV Fermilab Proton Driver beam from the SC 
linac would not be used to fill the MI, it would be available to drive a 
high intensity muon source. In addition to high intensity, precision muon 
experiments also require an appropriate bunch structure, which varies with 
experiment.In the post-Tevatron-Collider period it might be possible to 
utilize the Recycler Ring to 
repackage the 8~GeV proton beam, yielding a bunch structure optimized for each 
experiment. The combination of Proton Driver plus Recycler Ring would provide 
the front-end for a unique muon source with intensity and flexibility 
that exceed any existing facility.

\begin{table}[b] 
\caption{\label{tab:muon_sens} 
A comparison of the present or near future sensitivities for some 
representative muon experiments with the sensitivities that are in principle 
attainable with a Fermilab Proton Driver.
 } 
  \small 
  \begin{tabular}{||l|c|c||}\hline\hline 
  {} &{} &{} \\
Measurement & Present or Near Future & Fermilab Proton Driver \\
  {} &{} &{} \\
\hline
  {} &{} &{} \\
 EDM $d_\mu$    &  $< 3.7 \times 10^{-19}$ e-cm  & $< 10^{-24} - 10^{-26}$ e-cm \\
  $(g-2)$ $\sigma(a_\mu)$ & $0.2 - 0.5$ ppm      &      0.02 ppm                \\
  BR($\mu\rightarrow e\gamma$) &  $\sim 10^{-14}$ &      $\sim 10^{-16}$         \\
  $\mu A \rightarrow e A$ Ratio & $\sim 10^{-16}$ (?)    &      $\sim 10^{-19}$ \\ 
  {} &{} &{} \\
\hline\hline
\end{tabular}
\end{table}

The Recycler is an 8 GeV storage ring in the MI tunnel that can run at the same 
time as the MI. The beam from the Fermilab Proton Driver SC linac that is 
not used to fill the MI could be used to fill the Recycler Ring approximately ten 
times per second.  The ring would then be emptied gradually in the 100~ms intervals 
between linac pulses. Extraction could be continuous or in bursts.
For example, the Recycler Ring could be loaded with one linac pulse of 
$1.5 \times 10^{14}$ protons every 100~ms, with one missing pulse every 1.5 seconds 
for the 120 GeV MI program.  This provides $\sim 1.4 \times 10^{22}$ protons at 
8~GeV per operational year 
($10^7$~seconds). In the Recycler each pulse of $1.5 \times 10^{14}$ protons can 
be chopped into 588 bunches of $0.25 \times 10^{12}$ protons/bunch with a pulse 
width of 3~ns.  A fast kicker would permit the extraction of one bunch at a time. 
The beam structure made possible by the Proton Driver linac and the Recycler Ring 
is perfect for $\mu \rightarrow$ e conversion experiments, muon EDM searches and 
other muon experiments where a pulsed beam is required.  Slow extraction from the 
Recycler Ring for $\mu \rightarrow e \gamma$ and $\mu \rightarrow 3e$ searches is 
also possible. 

Using an 8~GeV primary proton beam together with a suitable target and 
solenoidal capture and decay channel, the calculated yield of low energy 
muons is $\sim 0.2$ of each sign per incident proton~\cite{brice}. 
With $1.4 \times 10^{22}$ 
protons at 8~GeV per operational year (corresponding to $\sim2$ megawatts) 
this would yield $\sim 3 \times 10^{21}$ muons per year. This muon flux greatly 
exceeds the flux required to make progress in a broad range of muon experiments. 
However, the muons at the end of the decay channel have low energy, a large momentum 
spread, and occupy a large transverse phase space. Without further manipulation their 
utilization will be very inefficient. The interface 
between the decay channel and each candidate experiment has yet to be designed. In 
Japan a Phase Rotated Intense Slow Muon Source 
(PRISM~\cite{prism}) based on an FFAG ring that 
reduces the muon energy spread (phase rotates) is being designed. This phase 
rotation ring has a very large transverse acceptance ($800\pi$~mm-mrad) and a momentum 
acceptance of $\pm30\%$ centered at 500~MeV/c. PRISM reduces the momentum and momentum 
spread to 68~MeV/c and $\pm 1-2\%$ respectively. Hence, a PRISM-like ring 
downstream of 
the decay channel might accept a significant fraction of the muon spectrum 
and provide a relatively  efficient way to use the available muon flux. Explicit 
design work must be done to verify this, but it should be noted that a muon 
selection system that utilizes only $1\%$ of the muons available at the end of the 
decay channel will still produce an adequate muon flux for most of the desired 
cutting-edge experiments.
Scaling from  proposals for muon experiments at JPARC, and making some plausible 
assumptions about the evolution of detector technology in the coming decade, 
the sensitivities that might be obtained at a Fermilab Proton Driver muon 
source are summarized in Table~\ref{tab:muon_sens} for the leading desired experiments.
Orders of magnitude improvements in sensitivity beyond those already 
achieved would be possible.

\section{Acknowledgments} 
The results summarized in this paper are from the Fermilab Proton Driver 
Physics Study, and therefore include contributions from all those that 
have participated. I am particularly indebted to the Working Group 
Conveners who organized and pushed forward the study: D.~Harris, 
S.~Brice, W.~Winter, J.~Morfin, R.~Ransome, R.~Tayloe, R.~Ray, 
L.~Roberts, H.~Nguyen, T.~Yamanaka, D.~Christian, M.~Mandlekern, 
H.~Cheung, P.~Kasper, P.~Ratoff, T.~Bowles, and G.~Green. 
This work was supported at Fermilab under contract DOE DE-AC02-76CH03000 
with the U.S. Department of Energy.

\end{document}